\documentclass[aps,prapplied,twocolumn,superscriptaddress,final]{revtex4-2}

\pdfoutput=1

\usepackage[colorlinks=true,allcolors=blue]{hyperref}
\usepackage{graphicx}
\usepackage{amsmath,amssymb,mathrsfs,array,longtable,delarray}
\usepackage{dcolumn}
\usepackage{bm}
\usepackage{textcomp} 

\begin{document}
\title{Pulsed Electroluminescence in a Dopant-free Gateable Semiconductor}

\author{S. R. Harrigan}
\address{Institute for Quantum Computing, University of Waterloo, Waterloo N2L 3G1, Canada}
\address{Department of Physics and Astronomy, University of Waterloo, Waterloo N2L 3G1, Canada}
\address{Waterloo Institute for Nanotechnology, University of Waterloo, Waterloo N2L 3G1, Canada}

\author{F. Sfigakis}
\email{Corresponding author: francois.sfigakis@uwaterloo.ca}
\address{Institute for Quantum Computing, University of Waterloo, Waterloo N2L 3G1, Canada}
\address{Department of Physics and Astronomy, University of Waterloo, Waterloo N2L 3G1, Canada}
\address{Waterloo Institute for Nanotechnology, University of Waterloo, Waterloo N2L 3G1, Canada}
\address{Department of Electrical and Computer Engineering, University of Waterloo, Waterloo N2L 3G1, Canada}
\address{Department of Chemistry, University of Waterloo, Waterloo N2L 3G1, Canada}

\author{L. Tian}
\address{Institute for Quantum Computing, University of Waterloo, Waterloo N2L 3G1, Canada}
\address{Department of Electrical and Computer Engineering, University of Waterloo, Waterloo N2L 3G1, Canada}

\author{N. Sherlekar}
\address{Institute for Quantum Computing, University of Waterloo, Waterloo N2L 3G1, Canada}
\address{Department of Physics and Astronomy, University of Waterloo, Waterloo N2L 3G1, Canada}

\author{B. Cunard}
\address{Institute for Quantum Computing, University of Waterloo, Waterloo N2L 3G1, Canada}
\address{Department of Electrical and Computer Engineering, University of Waterloo, Waterloo N2L 3G1, Canada}

\author{\\M. C. Tam}
\address{Waterloo Institute for Nanotechnology, University of Waterloo, Waterloo N2L 3G1, Canada}
\address{Department of Electrical and Computer Engineering, University of Waterloo, Waterloo N2L 3G1, Canada}

\author{H. S. Kim}
\address{Waterloo Institute for Nanotechnology, University of Waterloo, Waterloo N2L 3G1, Canada}
\address{Department of Electrical and Computer Engineering, University of Waterloo, Waterloo N2L 3G1, Canada}

\author{Z. R. Wasilewski}
\address{Institute for Quantum Computing, University of Waterloo, Waterloo N2L 3G1, Canada}
\address{Department of Physics and Astronomy, University of Waterloo, Waterloo N2L 3G1, Canada}
\address{Waterloo Institute for Nanotechnology, University of Waterloo, Waterloo N2L 3G1, Canada}
\address{Department of Electrical and Computer Engineering, University of Waterloo, Waterloo N2L 3G1, Canada}

\author{M. E. Reimer}
\email{mreimer@uwaterloo.ca}
\address{Institute for Quantum Computing, University of Waterloo, Waterloo N2L 3G1, Canada}
\address{Department of Physics and Astronomy, University of Waterloo, Waterloo N2L 3G1, Canada}
\address{Waterloo Institute for Nanotechnology, University of Waterloo, Waterloo N2L 3G1, Canada}
\address{Department of Electrical and Computer Engineering, University of Waterloo, Waterloo N2L 3G1, Canada}

\author{J. Baugh}
\email{baugh@uwaterloo.ca}
\address{Institute for Quantum Computing, University of Waterloo, Waterloo N2L 3G1, Canada}
\address{Department of Physics and Astronomy, University of Waterloo, Waterloo N2L 3G1, Canada}
\address{Waterloo Institute for Nanotechnology, University of Waterloo, Waterloo N2L 3G1, Canada}
\address{Department of Chemistry, University of Waterloo, Waterloo N2L 3G1, Canada}

\begin{abstract}
We report on a stable form of pulsed electroluminescence in a dopant-free direct bandgap semiconductor heterostructure which we coin the \emph{tidal effect}. Swapping an inducing gate voltage in an ambipolar field effect transistor allows incoming and outgoing carriers of opposite charge to meet and recombine radiatively. We develop a model to explain the carrier dynamics that underpins the frequency response of the pulsed electroluminescence intensity. Higher mobilities enable larger active emission areas than previous reports, as well as stable emission over long timescales.
\end{abstract}

\maketitle

\section{Introduction}

A pulsed form of electroluminescence (EL) has recently been discovered in ambipolar field-effect transistors (FETs) using direct bandgap 2D materials \cite{lien_large-area_2018}, which does not require a forward bias for light emission to occur. Instead, by periodically swapping the gate polarity of the FET, carriers already present in the FET recombine radiatively with incoming carriers of the opposite charge. Similar forms of this pulsed EL have been demonstrated in several material systems including 2D materials \cite{lien_large-area_2018,paur_electroluminescence_2019,zhao_generic_2020,zhu_highefficiency_2021},
bulk semiconductors \cite{hettick_shape-controlled_2020,cheng_hybrid_2020},
and Si CMOS \cite{rahman_low_2022,Tsang1997}. Pulsed EL is a promising candidate for light emitters with high wall-plug efficiency \footnote{Wall-plug efficiency is defined as the ratio of the total optical output power to the input electrical power drawn from the power grid.} since only a single ac signal is required for EL to be observed without the need for ac-to-dc conversion \cite{zak_bidirectional_nodate}.

Despite these demonstrations of pulsed EL, they suffer from the following disadvantages. First, all previous implementations have a single electrical contact, which allows efficient carrier injection for only one type of carrier polarity. Carrier injection is inefficient for the other carrier polarity due to the Schottky contact, thus limiting total brightness \cite{lien_large-area_2018}. In addition, they all have low mobilities which limit EL emission to small active areas, contained
within 20 \textmu m from the single electrical contact. For 2D materials, the EL intensity decays with time (minutes/hours) through permanent device degradation \cite{lien_large-area_2018,zhao_generic_2020} and device fabrication methods are not yet scalable in the semiconductor industry. Finally, bulk semiconductors and Si CMOS suffer from a low
emission efficiency. This inefficiency is due to a lack of confinement in bulk semiconductors, and an indirect bandgap in Si CMOS.

In this article, we overcome the above limitations by fabricating FETs in a III-V semiconductor heterostructure without doping, with efficient light emission and large active areas. Bright EL emission is observed up to 370 \textmu m away from the electrical contact, over an order of magnitude further away from the electrical contact than in any previous report \cite{lien_large-area_2018,paur_electroluminescence_2019,zhao_generic_2020,zhu_highefficiency_2021,hettick_shape-controlled_2020,cheng_hybrid_2020,rahman_low_2022} and limited only by the gate dimensions of our devices. Our implementation uses dedicated Ohmic contacts for both electrons and holes, which allows for rapid injection of both carrier types into the light emitting region. Neither EL decay nor device degradation are observed. Finally, temperature-dependent measurements allow the unambiguous identification of free excitons in the quantum well driving the observed EL.

\section{Experimental details and results}

\begin{figure}[t]
  \includegraphics[width=1\columnwidth]{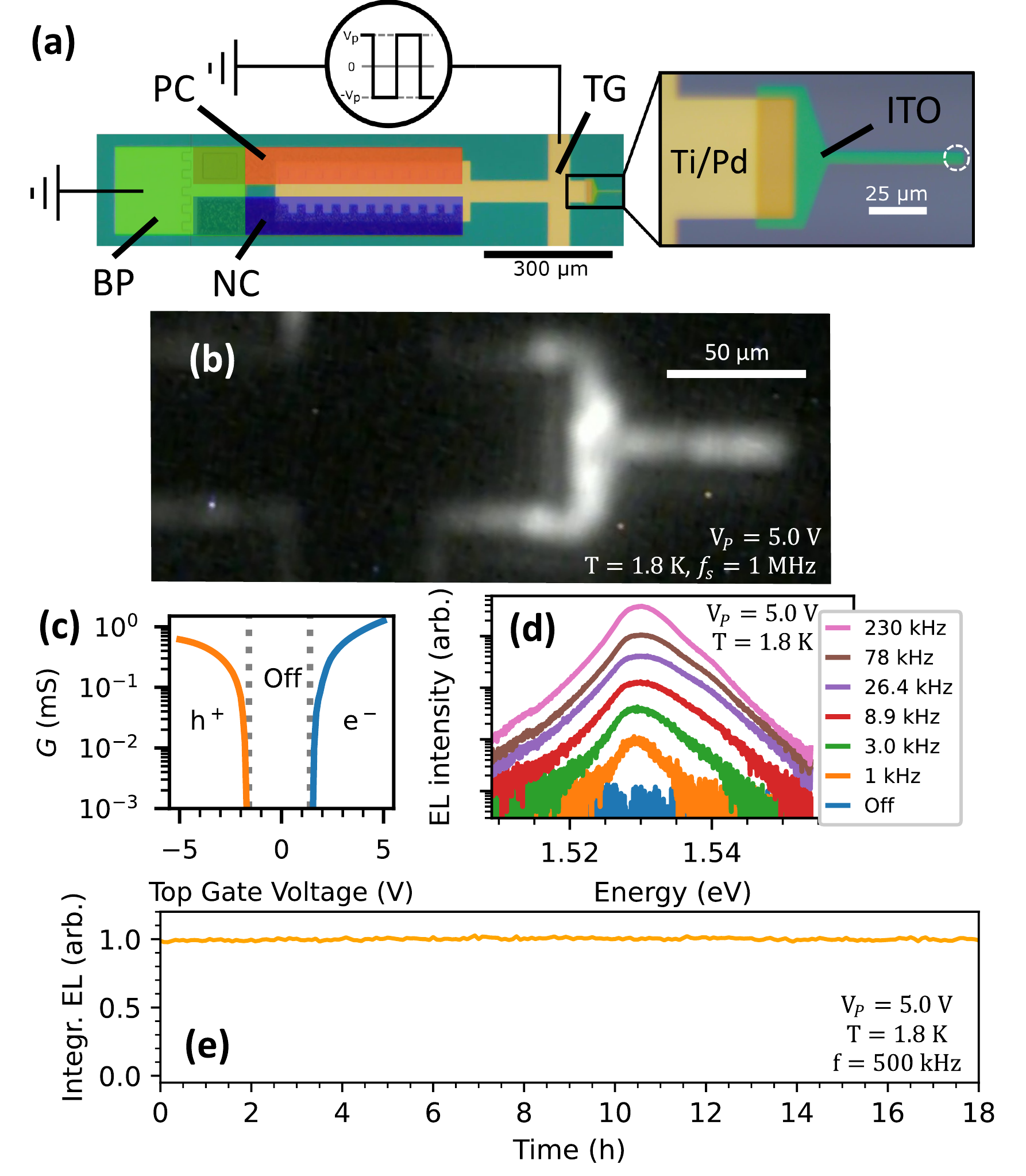}
  \caption{(a) Optical image of the sample and schematic of electrical circuit, with some device components false-colored for clarity. N-type and p-type Ohmic contacts are grounded during all measurements. A square wave with amplitude, $V_{p}$, and swapping frequency, $f_{s}$, is applied to the top gate. BP (green): Ohmic bond pad, NC (blue): n-type Ohmic contact, PC (orange): p-type Ohmic contact, TG (gold): top gate. (far right) Magnified view of device near the end of the top gate. Dashed white circle on the far right is the approximate collection area of the objective lens while acquiring EL spectra. Ti/Pd (gold): titanium/palladium, ITO (turquoise): indium tin oxide. (b) EL intensity map. EL cannot be observed through the opaque Ti/Pd gate, but faint EL around the edges of the Ti/Pd is observed. (c) Ambipolar electrical conductance, $G$, of the device, showing the turn-on of the FET for electrons, $\text{e}^{-}$, and holes $\text{h}^{+}$. (d) EL spectra generated from the tidal effect at varying swapping frequencies. (e) EL intensity of the tidal effect as a function of time.}
  \label{fig:sample_image}
\end{figure}

Dopant-free accumulation-mode GaAs-based FETs were fabricated from a 15 nm wide quantum well (QW) heterostructure grown by molecular beam epitaxy (MBE). Both n-type and p-type Ohmic contacts are present {[}Fig.~\ref{fig:sample_image}(a){]}, which allows the QW to be populated by either a two-dimensional electron gas (2DEG) or a two-dimensional hole gas (2DHG) in the same region underneath the top gate when a positive (negative) voltage beyond a threshold value is applied to
the top gate {[}Fig.~\ref{fig:sample_image}(c){]}. The two contacts are shorted together via a common bond pad, which allows them to effectively act as one ambipolar Ohmic contact \cite{chen_fabrication_2012,chung_quantized_2019,tian_stable_2023}. See the Appendix for additional details on MBE growth, sample fabrication, and measurement setups. The quantum well yielded an electron (hole) mobility of $7.68\times10^{5}$ cm$^{2}$ V$^{-1}$ s$^{-1}$ ($3.81\times10^{5}$ cm$^{2}$ V$^{-1}$ s$^{-1}$) at a carrier density of $2.69\times10^{11}$cm$^{-2}$ ($2.43\times10^{11}$cm$^{-2}$) at $T=1.6$ K in a dedicated Hall bar (see Fig.~\ref{fig:HallBar} in the Appendix). Unless otherwise specified, all measurements were performed at temperature $T=1.8$~K.

A typical device used for experiments is shown in Fig.~\ref{fig:sample_image}(a). Emitted light is only collected from the tip of the transparent indium tin oxide (ITO) top gate [dashed circle in inset of Fig.~\ref{fig:sample_image}(a)]. Figure \ref{fig:sample_image}(d) shows the EL spectrum of the device in Fig.~\ref{fig:sample_image}(a) during operation, with a corresponding image of light emission shown in Fig.~\ref{fig:sample_image}(b). The distance from the edge of the Ohmic contacts to the tip of the ITO top gate is 370~\textmu m. Figure \ref{fig:sample_image}(e) illustrates the time stability of the EL intensity during operation, with no sign of any decay over a duration of 18 hours. Furthermore, the same devices have been measured on and off over a period of one year, without degradation of EL properties.

\begin{figure*}[!th]
  \includegraphics[width=0.95\textwidth]{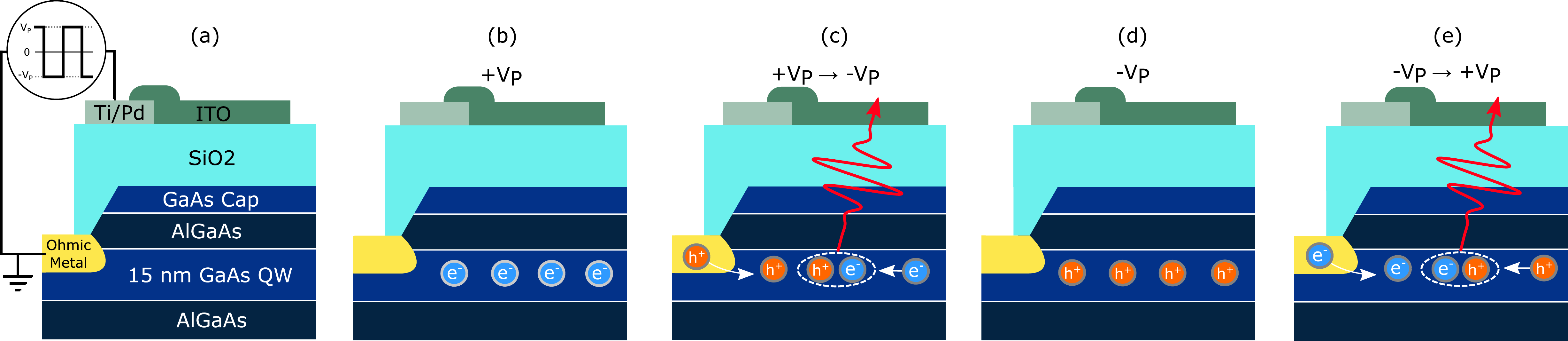}
  \caption{(a) Schematic view of device along growth direction (not to scale) indicating relevant layers within the device. (b)\textminus (e) Light emission mechanism of the tidal effect. (b) Initially, the top gate voltage is positive, and a 2DEG is induced in the QW. (c) The gate voltage is changed from positive to negative and holes are induced from the Ohmic contact and populate the QW. Simultaneously, the electrons retreat from the QW back to the Ohmic contact. Some electrons and holes meet while traveling and recombine radiatively, leading to a pulse of light emission. (d) After all the electrons have left the active area, a static 2DHG exists due to the negative gate voltage. (e) The gate voltage is changed from negative to positive, and a similar process to (c) occurs, except now with electrons (holes) entering (exiting) the QW.}
  \label{fig:schematic}
\end{figure*}

Figure \ref{fig:schematic} illustrates the mechanism for the observed pulsed luminescence. When swapping the polarity of the top gate voltage, the free carriers previously present under the top gate and furthest away from the Ohmic contact travel towards the Ohmic contact. Before reaching the Ohmic contact, they meet the incoming carriers of the opposite charge, and recombine radiatively. We call this the \emph{tidal effect} due to the incoming and outgoing flow of carriers during every top gate voltage swap. All data presented in the main text came from one sample, but the tidal effect was observed in all four samples, shown in Fig.~\ref{fig:Reproducibility} of the Appendix.

Figure \ref{fig:Frequency-and-voltage}(a) plots the integrated EL intensity as a function of peak top gate voltage $V_{p}$. As expected, brightness increases with increasing $V_{p}$, due to the higher carrier density. No EL occurs when $V_{p}<1.3$ V, in good agreement with the turn-on threshold voltage of the Ohmic contacts in Fig. \ref{fig:sample_image}(c) \footnote{In order to measure current in Fig. \ref{fig:sample_image}(b), both Ohmic contacts must be beyond their threshold voltage. To observe the tidal effect, however, only one Ohmic contact must be beyond its
threshold voltage. Therefore, it is possible for the onset of tidal effect light to occur at a lower voltage amplitude than the measured electrical turn on.}.

We studied the EL intensity as a function of gate swapping frequency, $f_{s}$. There are two regimes depending on the interplay between $\tau_{s}\equiv f_{s}^{-1}$ and the average carrier travel time, $\tau_{tr}$, from the Ohmic contact to the light collection area.

First, when $\tau_{s}\gg\tau_{tr}$, the EL intensity is expected to be linearly proportional to $f_{s}$ since all recombination has occurred well before the next gate voltage swap. Figure \ref{fig:Frequency-and-voltage}(b) plots the integrated peak intensity as a function of $f_{s}$ in this regime. A fit to a power law yields a best fit exponent of $m=0.954\pm0.004$, in excellent agreement with this hypothesis.

Second, when $\tau_{s} \lesssim \tau_{tr}$, not all carriers will arrive to the light collection area before the next gate voltage swap and, as a result, the EL intensity as a function of frequency is expected to saturate and then decline with increasing $f_{s}$. Figure \ref{fig:Frequency-and-voltage}(c) shows the EL frequency response in this regime for $0.2\leq f_{s}\leq5.7$ MHz.

As a simple model, let us treat the system in one dimension and consider the transient incoming carrier density, $n(x,t)$, where $x=0$ is the edge of the Ohmic contact and $t=0$ is the time when the top gate voltage is swapped. It is assumed that the incoming carrier density rapidly reaches its steady-state density, $N_{0}$, adjacent to the Ohmic contacts at $t=0$ [i.e.~$n(x\leq0,0)=N_{0}$] and that the area away from the Ohmic contacts is devoid of incoming carriers [i.e.~$n(x>0,0)=0$]. Thus, $n(x,t)$ will evolve according to the drift-diffusion transport equation in semiconductors:
\begin{equation}
\dfrac{\partial n}{\partial t}=-\mu E\dfrac{\partial n}{\partial x}+D\dfrac{\partial^{2}n,}{\partial x^{2}},\label{eq:drift_diffusion}
\end{equation}
\noindent where $\mu$ is the carrier mobility, $E$ is the electric field and $D$ is the diffusion constant. Performing a change of variables to a moving reference frame, $\xi=x-\mu Et$, we obtain $\partial n/\partial t=\partial^{2}n/\partial\xi^{2}$, which is Fick's second law in the moving frame. Solving for $\xi$ and substituting back to the stationary frame, we obtain:
\begin{equation}
n(x,t)=\frac{N_{0}}{2}\text{erfc}\Bigg(\frac{x-\mu Et}{2\sqrt{Dt}}\Bigg),\label{eq:transient_carrier_density}
\end{equation}
\noindent where $\text{erfc}(x)\equiv1-\text{erf}(x)$. EL can only occur when both carrier polarities are present, which occurs when the first wave of incoming carriers arrives. Since the EL will be limited by the rate of incoming carriers, the total EL intensity as a function of swapping frequency, $I\left(f_{s}\right)$, will be proportional to $f_{s}$ multiplied by $\left|\partial n/\partial t\right|$ integrated over the swapping period $\tau_{s}$:
\begin{eqnarray}
I\left(f_{s}\right) &=& cf_{s}\int_{0}^{\tau_{s}}\left|\partial n/\partial t\right|dt\nonumber\\
&=&\frac{cf_{s}N_{0}}{2}\,\textnormal{erfc}\Bigg(\frac{xf_{s}-\mu E}{2\sqrt{Df_{s}}}\Bigg), \label{eq:EL_v_f_model}
\end{eqnarray}
\noindent where $c$ is a scaling factor accounting for efficiencies of intrinsic (e.g. radiative recombination probability) and extrinsic (e.g. collection and detector efficiencies) nature.

\begin{figure}[!b]
  \includegraphics[width=0.95\columnwidth]{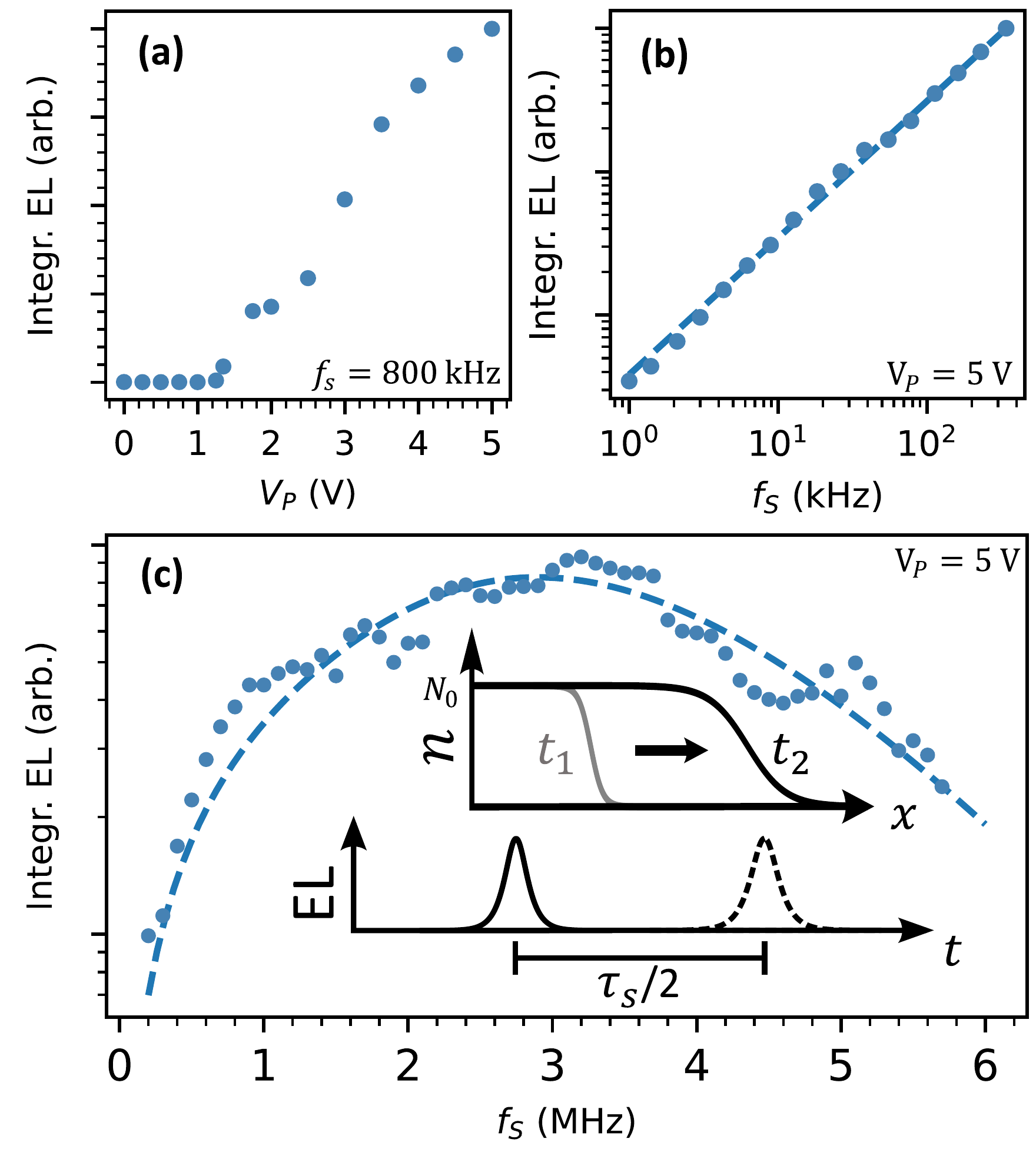}
  \caption{(a) Integrated EL intensity as a function of peak voltage amplitude $V_{P}$. (b) Integrated EL intensity as function of swapping frequency in the low frequency regime ($\tau_{s}\gg\tau_{tr}$). Dashed line is a fit to a power law. (c) Integrated EL intensity as a function of swapping frequency in the high frequency regime ($\tau_{s} \lesssim \tau_{tr}$). Dashed line is a fit to Eq.~(\ref{eq:EL_v_f_model}). (top inset) Schematic representation of carrier density $n(x,t)$ while traveling. With increasing time, the carrier profile shifts to the right (due to drift) and spreads out (due to diffusion). (bottom inset) Diagram of pulsed EL as a function of time at a fixed position. When swapping the gate voltage polarity at $f_{s}$, the next EL pulse will occur after time $\tau_{s}/2$.}
  \label{fig:Frequency-and-voltage}
\end{figure}

In our case, EL is observed at a constant distance from the Ohmic contact ($x=370$ $\mu$m). The voltage difference across the sample is estimated to be $\sim$\,1.5~V at $t=0$, as the quasi-Fermi levels span from just above the conduction band to just below the valence band across the 370~\textmu m distance, and therefore, $\left|E\right|\sim$\,40.5 V cm$^{-1}$. Equation~(\ref{eq:EL_v_f_model}) was fit to data shown in Fig.~\ref{fig:Frequency-and-voltage}(c) with $\mu$, $D$, and $c$ left as fitting parameters. We obtain $\mu=3.55\pm0.08\times10^{3}$ cm$^{2}$ V$^{-1}$ s$^{-1}$ and $D=290\pm40$ cm$^{2}$ s$^{-1}$, consistent with the values for bulk GaAs in small electric fields \cite{ruch_transport_1968}, with $\mu$ approximately 2 orders of magnitude lower than in our reference Hall bar. At low carrier densities in our samples, Thomas-Fermi screening of charged impurities is the dominant mechanism for increasing mobility \cite{ando_electronic_1982,shetty_effects_2022}. The first wave of incoming carriers experiences no screening and, as a result, the carrier mobility is close to that of bulk GaAs.

\begin{figure}[t]
\includegraphics[width=1\columnwidth]{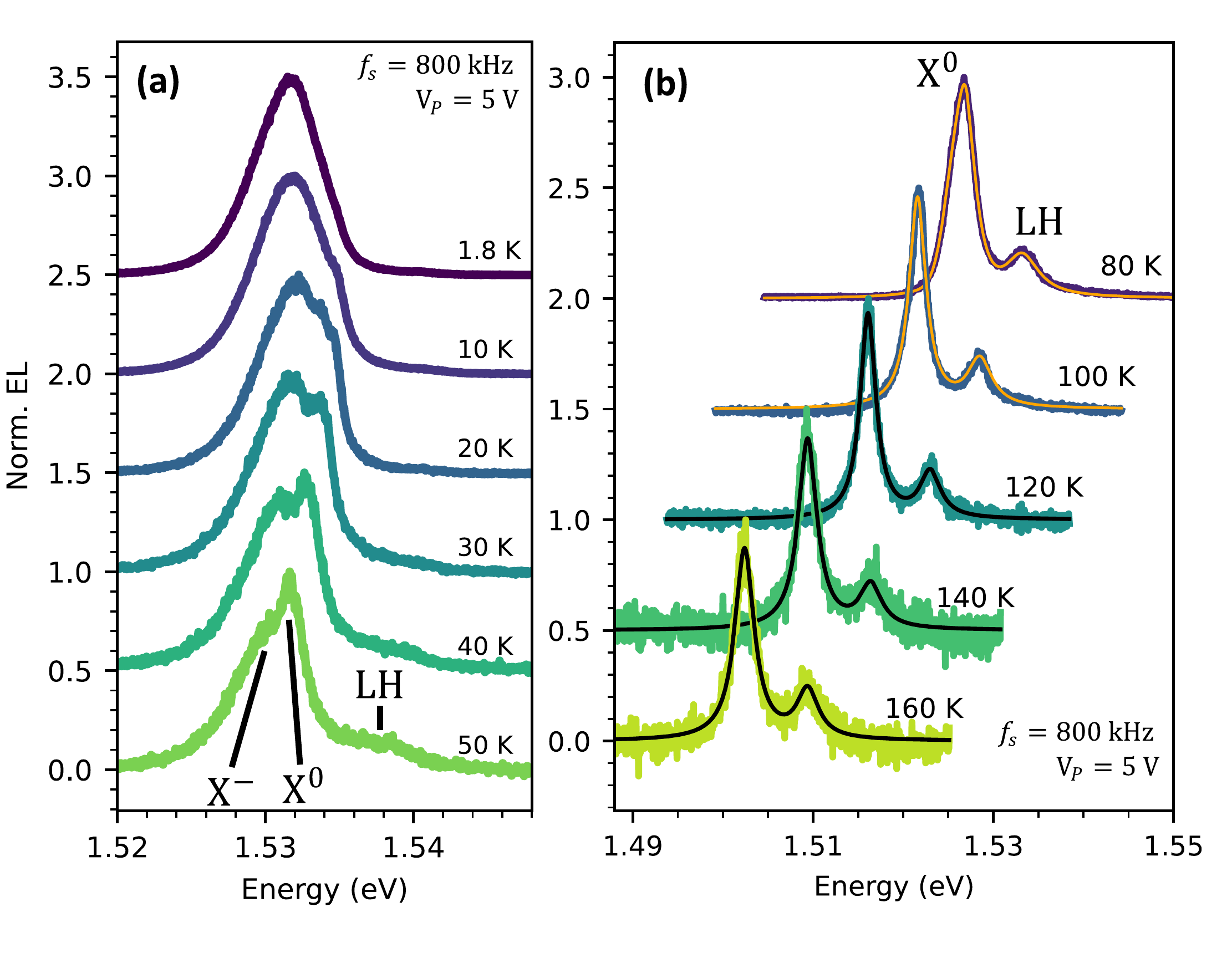}
  \caption{Temperature dependent spectra at: (a) low temperatures and (b) high temperatures, with traces offset for clarity. See main text for peak label definitions. Solid lines in panel (b) are best fits using two Lorentzians peaks.}
  \label{fig:Temperature-dependece}
\end{figure}

To identify the EL emission peak in Fig.~\ref{fig:sample_image}(d) and show that the tidal effect persists at elevated temperatures, we perform EL measurements from $T=1.8$~K to 160~K, shown in Fig.~\ref{fig:Temperature-dependece}. At $T=1.8$~K, the EL emission spectrum is dominated by a peak centered at 1.531~eV with an asymmetric low-energy tail [see Fig.~\ref{fig:Temperature-dependece}(a)]. This is consistent with the negatively charged exciton, $\text{X}^{-}$\cite{tian_stable_2023}, where the low-energy tail is due to an electron recoil effect \cite{esser_photoluminescence_2000,ross_electrical_2013,christopher_long_2017}. With increasing temperature, two higher-energy peaks become progressively more prominent. The first matches the expected emission energy ($\sim$1.534~eV at $T=1.8$~K) for the heavy hole neutral exciton (X$^{0}$) in a 15 nm GaAs QW \cite{hsiao_single-photon_2020,tian_stable_2023}. The second, a shallow peak at even higher energy, is attributed to the light hole neutral exciton (LH) \cite{shields_magneto-optical_1995,tian_stable_2023}. For $T>80$~K [see Fig.~\ref{fig:Temperature-dependece}(b)], X$^{-}$ disappears and the two remaining peaks are attributed to X$^{0}$ and LH \cite{tian_stable_2023}. The energy splitting between the two peaks remains roughly constant from 90~K to 160~K at (6.93$\pm$0.09) meV, consistent with this assignment \cite{tian_stable_2023}. Both peaks fit well to a Lorentzian lineshape, implying that X$^{0}$ and LH are experiencing homogeneous broadening. This rules out charge noise (a common inhomogeneous broadening mechanism) as a significant source of linewidth broadening in our devices, which is expected from our high-quality dopant-free material. With increasing temperature, LH becomes more prominent relative to X$^{0}$, as holes transfer from the HH band to the LH band \cite{shields_magneto-optical_1995,lee_luminescence_1986}. In addition, the total EL intensity decreases with increasing temperature, as routinely observed in PL experiments in GaAs QWs \cite{jiang_temperature_1988,chiari_temperature_1988,lambkin_thermal_1990,dang_photoluminescence_2009},
and is not observed at temperatures beyond 160~K. We attribute the reduction in EL due to the higher probability of carriers tunneling out of the QW as the temperature increases \cite{lambkin_thermal_1990}.

\section{Discussion and Conclusions}

In order to achieve higher-temperature operation, it is necessary to increase the conduction band offset, $\Delta E_{c}$, and valence band offset, $\Delta E_{v}$, between the QW and the barrier material. In this study, we used an Al$_{0.3}$Ga$_{0.7}$As barrier, with $\Delta E_{c}=217$ meV and $\Delta E_{v}=$164 meV to the GaAs QW \cite{tian_stable_2023}. Using Al$_{0.45}$Ga$_{0.55}$As barriers will result in $\Delta E_{c,v}\approx250$ meV \cite{batey_energy_1986,wang_band_2013}. Alternatively, a GaAs QW with In$_{0.5}$Ga$_{0.15}$Al$_{0.35}$P barriers will result in $\Delta E_{c}=380$ meV and $\Delta E_{V}\approx500$ meV \cite{watanabe_interface_1987}. Since the probability of a carrier tunneling out of the QW is exponential in $\Delta E_{c/v}$, values of $\Delta E_{c/v}\approx400$ meV should be sufficient for tunneling to be negligible at room temperature over the timescales required for the incoming carriers to populate the QW \cite{lambkin_thermal_1990}. This is consistent with room temperature observation of the tidal effect in 2D semiconductors on SiO$_{2}$ \cite{lien_large-area_2018}, where $\Delta E_{c/v}>1$ eV.

In summary, we have demonstrated pulsed EL in III-V FET devices driven by the tidal effect for the first time. Taking into account the transient dynamics of the incoming carriers, we developed a model that describes the integrated EL intensity as a function of swapping frequency, which allowed us to determine the effective carrier mobility and diffusion constant in our devices. Due to faster travel times, we observed EL over longer distances from the Ohmic contacts by more than an order of magnitude as compared to previous reports. We also found that the EL originated from electron-hole recombination of free excitons and trions in the QW. Furthermore, our implementation uniquely used dedicated Ohmic contacts for both electrons and holes, allowing for efficient carrier injection and brighter EL as a result.

\begin{acknowledgments}
S.R.H., F.S., and L.T. contributed equally. F.S., M.E.R., and J.B. supervised the work equally. We thank members of the Quantum Photonic Devices Lab for valuable discussions. This research was undertaken thanks in part to funding from the Canada First Research Excellence Fund (Transformative Quantum Technologies), Defence Research and Development Canada (DRDC), and Canada\textquoteright s Natural Sciences and Engineering Research Council (NSERC). S.R.H. acknowledges further support from a NSERC Canada Graduate Scholarships-Doctoral (CGS-D). The University of Waterloo's QNFCF facility was used for this work. This infrastructure would not be possible without the significant contributions of CFREF-TQT, CFI, ISED, the Ontario Ministry of Research \& Innovation and Mike \& Ophelia Lazaridis. Their support is gratefully acknowledged.
\end{acknowledgments}

\section*{Data availability}

The data that support the findings of this paper are openly available from the Zenodo archive \cite{zenodo-tidal}.

\appendix*
\section{Additional experimental details}

\begin{figure*}[t]
  \includegraphics[width=2.05\columnwidth]{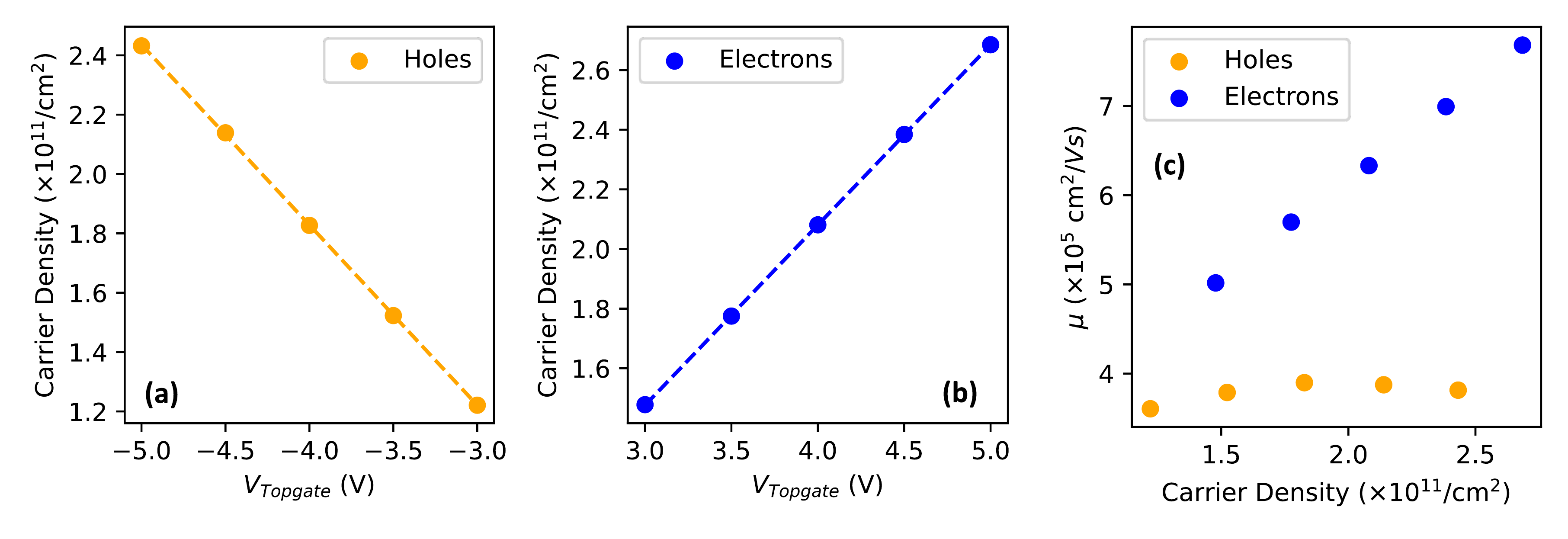}
  \caption{(a) Hole density and (b) electron density as a function of top gate voltage. Dashed lines are linear fits. (c) Electron and hole mobility as a function of carrier density.}
  \label{fig:HallBar}
\end{figure*}

\textbf{MBE growth.} The quantum well heterostructure used in this experiment, wafer G377, was grown by molecular beam epitaxy on a semi-insulating GaAs (100) substrate. Starting from the substrate, the layer order is as follows: a 200 nm GaAs buffer, a 20-period smoothing superlattice composed of a of 2.5 nm GaAs and 2.5 nm Al$_{0.3}$Ga$_{0.7}$As layer, a 500 nm Al$_{0.3}$Ga$_{0.7}$As barrier, a 15 nm wide GaAs quantum well, a 150 nm Al$_{0.3}$Ga$_{0.7}$As barrier, and a 10 nm GaAs cap layer. No intentional dopants were placed in the heterostructure.

\begin{figure}[b]
  \includegraphics[width=1.00\columnwidth]{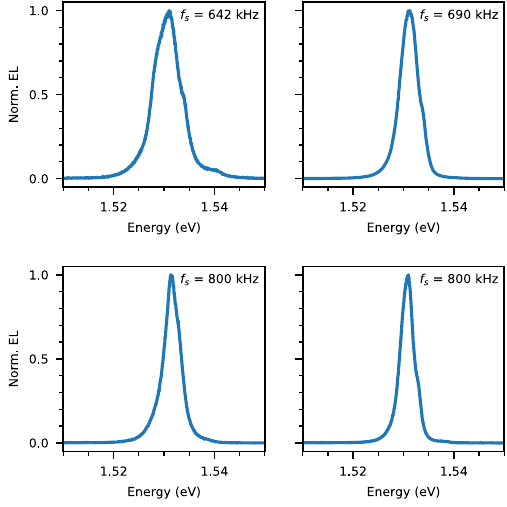}
  \caption{EL spectrum for 4 different samples, showing reproducibility of the tidal effect. In all 4 plots, a square wave with voltage amplitude $V_{p}$ = 5 V is applied to the top gate and $T = 1.8$ K.}
  \label{fig:Reproducibility}
\end{figure}

\textbf{Sample fabrication.} A mesa is defined by a wet etch in 1:8:120 H$_{2}$SO$_{4}$:H$_{2}$O$_{2}$:H$_{2}$O mixture. AuBe p-type and Ni/AuGe/Ni n-type recessed ohmic contacts were deposited and annealed at 520$^{\text{\textopenbullet}}$C for 180 seconds and 450$^{\text{\textopenbullet}}$C for 180 seconds, respectively. A 300 nm thick SiO$_{2}$ insulator layer was deposited by plasma-enhanced chemical vapor deposition. A 20/60 nm Ti/Pd top gate is deposited on top of the SiO$_{2}$ (overlapping with the ohmic contacts), and a 30 nm layer of indium tin oxide is deposited near the tip of the device.

\textbf{Carrier density and mobility characterization.} A dedicated Hall bar device was fabricated in tandem with the EL samples. It was used to assess the carrier density as a function of top gate voltage and mobility as a function of carrier density at $T = 1.6$ K, as shown in Figure~\ref{fig:HallBar}. The carrier density versus top gate voltage is linear and shows no hysteresis in the range measured ($\pm$5 V). Electron and hole mobilities are similar to those previously reported for quantum wells in undoped GaAs with similar well width \cite{tian_stable_2023}. We note that significantly higher electron mobilities ($>5\times10^{6}$ cm$^{2}$/V$\cdot$s) at similar densities has been demonstrated in undoped GaAs heterostructures by us \cite{shetty_effects_2022} and others \cite{lilly_resistivity_2003,pan_impact_2011,peters_gating_2016}.

\textbf{Measurement setups.} All optical measurements were performed in a Attocube AttoDRY 2100 optical cryostat with a base temperature of 1.6 K. Signals were generated using a Tektronix AFG1062 (rise time $<$ 10 ns per volt) and sent to the sample using home-built electrical feedthroughs. Light generated from the sample is collected using an Attocube LT-APO/NIR/0.81 objective lens and coupled in free-space to a Princeton SP-2-750i spectrometer equipped with a PIXIS: 100BR\_eXcelon CCD, using a groove density of 1200 gratings/mm.

\textbf{Reproducibility in many samples.} Figure \ref{fig:Reproducibility} shows EL spectra acquired with 4 different samples in identical experimental conditions to demonstrate the tidal effect is reproducible.

\begin{figure}[t]
  \includegraphics[width=0.90\columnwidth]{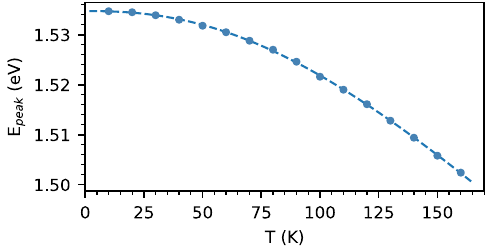}
  \caption{EL emission energy of HH X$^{0}$ as a function of temperature. The dashed line is a fit to the P\"{a}ssler model \cite{passler_basic_1997,lourenco_temperature_2001}.}
  \label{fig:Passler}
\end{figure}

\textbf{Temperature dependence of HH X$^0$.} Figure~\ref{fig:Passler} tracks HH X$^{0}$ as a function of temperature. From $T = 10$ K to $T = 160$ K, the emission peak red shifts by nearly 33 meV, primarily due to the decrease in the band gap energy of GaAs with increasing temperature \cite{pierret_advanced_2003}. The dashed line is a fit to the P\"{a}ssler model \cite{passler_basic_1997} for the temperature dependence of band gap energy. The HH X$^{0}$ binding energy is found to be $E_{b}=9.2\pm0.1$ meV, consistent with prior PL and EL measurements \cite{filinov_influence_2004,tian_stable_2023}.

\begin{figure}[h]
  \includegraphics[width=0.90\columnwidth]{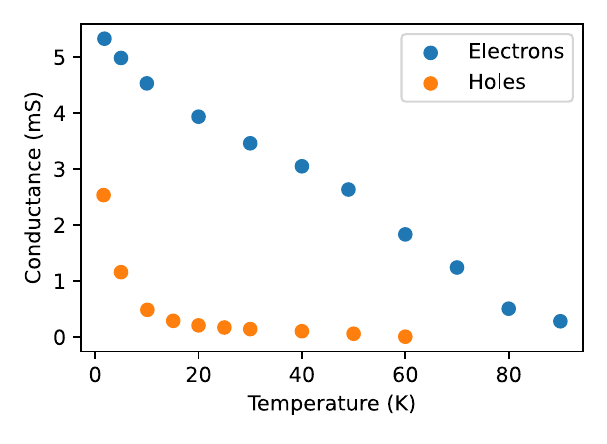}
  \caption{Constant-voltage four-terminal conductance of reference Hall bar as a function of temperature.}
  \label{fig:Tdep-HB}
\end{figure}

\textbf{Temperature dependence of Hall bar.} Measurements on the reference Hall bar indicate the 2DEG (2DHG) loses confinement at $T=90$ (65) K, as shown in Figure \ref{fig:Tdep-HB}. The lower operating temperature of the 2DHG is consistent with smaller band offset in the valence band ($\sim164$ meV) compared to the conduction band ($\sim217$ meV) \cite{tian_stable_2023}. However, these temperatures do not tally with the temperature at which the tidal effect quenches (160 K). Nevertheless, from the EL emission energy (1.534~eV), the radiative recombination unambiguously occurs within the 15 nm wide GaAs quantum well, and not in the surrounding AlGaAs barriers or GaAs cap layer. At high temperature, we speculate that pulsed transient carrier populations are channelled through the quantum well, for both incoming and outgoing carriers during each swapping cycle.

\newpage

\end{document}